\documentclass[a4paper,12pt, epsfig]{article}
\pdfoutput=1
\usepackage{epsfig}
\usepackage{epstopdf}
\usepackage{graphicx}
\usepackage{ifthen}

\usepackage{amsmath}
\usepackage[psamsfonts]{amssymb}
\usepackage{euscript}

\usepackage{latexsym}
\usepackage[arrow,matrix,curve]{xy}

\newskip\humongous \humongous=0pt plus 1000pt minus 1000pt

\newif\ifdtup

\def\,{\hspace{-.1cm}}
\def\hsp{,\hspace{.7cm}}

\def\bra#1{\langle#1|}
\def\ket#1{|#1\rangle}
\def\braket#1#2{\langle#1|#2\rangle}

\def\fc#1#2 {\frac{n}{q}#1\frac{n}{q}#2}

\renewcommand{\theequation}{\arabic{section}.\arabic{equation}}
\renewcommand{\(}{\begin{equation}}
\renewcommand{\)}{end{equation} \vspace{-.05in}\linebreak}

\newcounter{saveeqn}
\newcounter{savealpheqn}

\newcommand{\alpheqn}{\setcounter{saveeqn}{\value{equation}}%
  \stepcounter{saveeqn}\setcounter{equation}{0}%
  \renewcommand{\theequation}{\mbox{\arabic{section}.\arabic{saveeqn}
\alph{equation}}}
  \renewcommand{\)}{\end{equation}}}
\def\part#1{\frac{\partial}{\partial{#1}}}%
\def\group#1{\refstepcounter{equation}\setcounter{saveeqn}
 {\value{equation}}%
  \label{#1}\setcounter{equation}{0}%
\renewcommand{\theequation}{\mbox{\arabic{section}.\arabic{saveeqn}
\alph{equation}}}
  \renewcommand{\)}{\end{equation}}}
\newcommand{\reseteqn}{\setcounter{equation}{\value{saveeqn}}%
  \renewcommand{\theequation}{\arabic{section}.\arabic{equation}}%
  \renewcommand{\)}{\end{equation}}}

\newcommand{\aalpheqn}{\setcounter{saveeqn}{\value{equation}}%
  \stepcounter{saveeqn}\setcounter{equation}{0}%
  \renewcommand{\theequation}{\mbox{
        \Alph{subsection}.\arabic{saveeqn}\alph{equation}}}
   \renewcommand{\)}{\end{equation}}}
\newcommand{\areseteqn}{\setcounter{equation}{\value{saveeqn}}%
  \renewcommand{\theequation}{\Alph{subsection}.\arabic{equation}}%
  \renewcommand{\)}{\end{equation}}}

\renewcommand{\thefootnote}{\alph{footnote}}
\renewcommand{\(}{\begin{equation}}
\renewcommand{\)}{\end{equation}}
\newcommand{\ba}{\begin{eqnarray}}
\newcommand{\ea}{\end{eqnarray}}

\newcommand{\bp}{\mathop{\vtop{\ialign{##\crcr
   $\hfil\displaystyle{}\hfil$\crcr\noalign{\kern-13pt\nointerlineskip}
   \BIG{(}\hskip0pt\crcr\noalign{\kern3pt}}}}}
\newcommand{\cbp}{\mathop{\vtop{\ialign{##\crcr
   $\hfil\displaystyle{}\hfil$\crcr\noalign{\kern-13pt\nointerlineskip}
   \BIG{)}\hskip0pt\crcr\noalign{\kern3pt}}}}}
\newcommand{\pa}{\mathop{\vtop{\ialign{##\crcr
    
$\hfil\displaystyle{\oplus}\hfil$\crcr\noalign{\kern+1pt\nointerlineskip 
}
   \hspace{.08in}$^{\alpha=0}$\hskip6pt\crcr\noalign{\kern3pt}}}}}
\renewcommand{\hsp}{,\hspace{.3in}}
\newcommand{\p}{^\prime}

\catcode`\@=11
\def\vereq#1#2{\lower3pt\vbox{\baselineskip1.5pt \lineskip1.5pt
\ialign{$\m@th#1\hfill##\hfil$\crcr#2\crcr\sim\crcr}}}
\catcode`\@=12

\renewcommand{\(}{\begin{equation}}
\renewcommand{\)}{\end{equation}}

\newcommand{\beas}{\begin{eqnarray*}}
\newcommand{\eeas}{\end{eqnarray*}}

\newcommand{\bquo}{\begin{quote}}
\newcommand{\enqu}{\end{quote}}

\newcommand{\beq}{\begin{equation}}
\newcommand{\eeq}{\end{equation}}
\newcommand{\bea}{\begin{eqnarray}}
\newcommand{\eea}{\end{eqnarray}}

\newskip\humongous \humongous=0pt plus 1000pt minus 1000pt

\newif\ifdtup

\catcode`\@=11

\@addtoreset{equation}{section}

\def\@normalsize{\@setsize\normalsize{15pt}\xiipt\@xiipt
\abovedisplayskip 14pt plus3pt minus3pt%
\belowdisplayskip \abovedisplayskip
\abovedisplayshortskip \z@ plus3pt%
\belowdisplayshortskip 7pt plus3.5pt minus0pt}

\def\small{\@setsize\small{13.6pt}\xipt\@xipt
\abovedisplayskip 13pt plus3pt minus3pt%
\belowdisplayskip \abovedisplayskip
\abovedisplayshortskip \z@ plus3pt%
\belowdisplayshortskip 7pt plus3.5pt minus0pt
\def\@listi{\parsep 4.5pt plus 2pt minus 1pt
      \itemsep \parsep
      \topsep 9pt plus 3pt minus 3pt}}

\relax

\catcode`@=12

\catcode`\@=11

\def\section{\@startsection{section}{1}{\z@}{3.5ex plus 1ex minus  .2ex}{2.3ex plus .2ex}{\large\bf}}

\def\thesection{\arabic{section}}
\def\thesubsection{\arabic{section}.\arabic{subsection}}

\def\appendix{\setcounter{section}{0}
 \def\thesection{Appendix \Alph{section}}
 \def\thesubsection{\Alph{section}.\arabic{subsection}}
 \def\theequation{\Alph{section}.\arabic{equation}}}
\renewcommand{\theequation}{\arabic{section}.\arabic{equation}}

\begin{document}
\def\thefootnote{\fnsymbol{footnote}}
\def\thetitle{Ground States of the $\phi^4$ Double-Well QFT}
\def\autone{Hui Liu${}^{1,2}$}
\def\auttwo{Yao Zhou${}^{1,2}$}
\def\autthree{Jarah Evslin${}^{1,2}$}
\def\affa{Institute of Modern Physics, Chinese Academy of Sciences, Lanzhou 730000, China}
\def\affb{University of the Chinese Academy of Sciences, Beijing 100049, China}

\begin{center}
{\large {\bf \thetitle}}

\bigskip

\bigskip

{\large \noindent  \autone \footnote{liuhui@impcas.ac.cn}, \auttwo \footnote{yaozhou@impcas.ac.cn} and \autthree \footnote{jarah@impcas.ac.cn} }

\vskip.7cm

1) \affa\\
2) \affb\\

\end{center}

\begin{abstract}
\noindent

\noindent
At second order in perturbation theory, we find the ground states of the $\phi^4$ double-well quantum field theory in 1+1 dimensions.  The operators which create these ground states from the free vacuum are constructed explicitly at this order, as is the operator which interpolates between the ground states.  As a warmup we perform the analogous calculation in quantum mechanics, where the true ground state is unique but in perturbation theory there are also two ground states.


\end{abstract}

%
\setcounter{footnote}{0}
\renewcommand{\thefootnote}{\arabic{footnote}}




\section{Introduction}

One of the greatest triumphs in nonperturbative quantum field theory was the calculation of the leading order correction to the $\phi^4$ kink mass in Ref.~\cite{dhn2}.   This theory was regularized by placing it in a periodic box and fixing the number of modes considered.   The calculation was so successfully that it was performed again by a number of groups, many of whom found different results.  In Ref.~\cite{rv97} it was explained that the reason for this discrepancy is that the leading correction depends on the regularization scheme, with a fixed mode number and sharp energy cutoff leading to distinct results.  Since then even more schemes have been proposed, such as demanding that the correction be independent of the topological sector when the mass vanishes \cite{nastase}.  While various consistency checks have been performed, it remains unclear which scheme, if any, is correct or whether perhaps all are correct but with different definitions of the bare theory or renormalization conditions.

Almost immediately it had been noted \cite{raja75} that the regularization and renormalization of the $\phi^4$ theory were not necessary at all, as scalar theories in 2d with canonical kinetic terms are manifestly finite when normal-ordered, because the only divergences arise from loops involving a single vertex.  A manifestly finite calculation was therefore possible.  Such a calculation has recently appeared \cite{mekink}.  From the point of view of the original theory, one might argue that normal ordering is just another scheme.  We will take a different point of view, that the normal-ordered theory is a finite quantum field theory with a kink solution in which quantum corrections can be calculated free from any ambiguities.  Therefore the normal-ordered theory will be the starting point of our investigation.

Now that all ambiguities have been removed, one can for the first time meaningfully press deeper into the quantum regime.  It is now possible to unambiguously calculate the subleading corrections to the kink mass and indeed to the kink state itself.  However, to do this, one must first calculate the $\phi^4$ vacua and their energies to the same order in perturbation theory.  This is the goal of the present paper.

This is a straightforward exercise in perturbation theory.  But it has so far escaped attention\footnote{In the case of the single well $\phi^4$ theory, operators that create the ground state from the disentangled vacuum have recently been constructed in Refs.~\cite{ali1,ali2}.   The theory and renormalization scheme are quite different from ours, for example there is no $\phi^3$ potential in the single well, however their quartic term is similar to our corresponding result.} for several reasons.  First of all, the perturbation theory does not converge \cite{jaffe65}.  Therefore to describe such models rigorously, one convolves the Hamiltonian with a spatial cutoff \cite{gj1}.  However were one to include such a cutoff in the double-well theory, it would imply that beyond the cutoff there is a unique minimum of the potential.  One might still introduce two states, one with an expectation value of the scalar field near each minimum within the cutoff.  However, there would be finite action instantons which mix these states and the true ground state would be a unique, symmetric combination\footnote{Of course the usual logic is to begin by assuming the existence of a unique vacuum.}.  Therefore there is no stable kink in this theory.

We will ignore these problems, because our goal is eventually to understand the supersymmetric theory where the divergent vacuum energy at the heart of these problems does not appear.  Therefore we will proceed with a perturbative approach with no spatial cutoff.

We begin in Sec.~\ref{qm} with a description of the corresponding problem in quantum mechanics.  Here, just like in the position cutoff quantum field theory, there is a unique ground state \cite{cohen}.  However, at any order in perturbation theory there are in fact two distinct ground states, one localized in each well.  We will calculate these ground states and the operators which create them from the harmonic oscillator ground state.  Then in Sec.~\ref{qft} we perform the full computation in quantum field theory, calculating the corrections to the states and energy at second order.  We see that the solution is a straightforward generalization of that in quantum mechanics.  In Sec.~\ref{interp} we construct the operator which interpolates between the two ground states.  This operator is, in a sense, the $x_0\rightarrow \infty$ limit of the operator that creates a kink centered at $x_0$.

\section{The Double Well In Quantum Mechanics} \label{qm}

The double-well model in quantum mechanics has been solved many times, exactly and approximately.  In this section we will review the construction of the ground state(s) in second order perturbation theory, as a warm-up for the very similar calculation in quantum field theory in Sec.~\ref{qft}.  There are several reasons not to solve this model in perturbation theory.  One problem is that the perturbation theory does not converge \cite{bw}.  Of course this is a standard problem in quantum theories, where one is used to having only asymptotic expansions.  However in the present case it means that one can expect significant errors in the wave function whenever one is not close to the spatial minimum of the potential about which one defines the expansion.   

A more serious problem is that while the Schrodinger equation with no extra degrees of freedom always has a unique ground state solution \cite{cohen} this is not necessarily true in   perturbation theory.  In the present case, in perturbation theory there are in fact two ground states, one localized at each minimum of the well.  In the full, nonperturbative theory these two states are mixed by instantons with only the symmetric combination being the ground state.  However no instantons appear in perturbation theory.  This problem is actually an advantage for us, as it mimics the case in quantum field theory, where any such instantons would have infinite action and so there really are two distinct vacua.  Recall that in 1+1 dimensions global symmetries, such as our reflection symmetry, can be spontaneously broken if they are discrete.

We consider a particle in a double-well potential. The corresponding Hamiltonian is
\beq
H=\frac{p^2}{2m}+k \left(x^2-v^2\right)^2,
\eeq
where $m$ is the particle's mass, $k$ and $v$ are positive constants and we have set $\hbar=1$. If the well-separation $2v$ is sufficient, then in perturbation theory there will be two ground states, centered approximately on $x=v$ and $x=-v$, both with the same energy $E$.  We will denote these two states by $\ket{+v}$ and $\ket{-v}$
\beq
H\ket{\pm v}=E\ket{\pm v}.
\eeq
Our task is to find $E$ and also the wave functions of the two ground states.

To find the state $\ket{+v}$, we can rewrite the Hamiltonian as an anharmonic oscillator centered on $x=v$ 
\beq
H=\frac{p^2}{2m}+k \left((x-v)^4+4 v (x-v)^3+4 v^2 (x-v)^2\right).
\eeq
This can be simplified using the spatial displacement operator\beq
\mathcal{D}_{\alpha}={\rm{exp}}(-i \alpha p)
\eeq
to define a new Hamiltonian $H\p$ by
\beq
H\p=\mathcal{D}_{-v}H \mathcal{D}_{-v}^{-1}=\frac{p^2}{2m}+4kv^2 \left(x^2+\frac{1}{v}x^3+\frac{1}{4v^2} x^4\right),\label{qmh'}
\eeq
which again has two ground states.  We will be interested in that localized at $x=0$
\beq
\ket{g_1}=\mathcal{D}_{-v}\ket{+v}\hsp
H\p\ket{g_1}=\mathcal{D}_{-v}H \mathcal{D}_{-v}^{-1}\mathcal{D}_{-v}\ket{+v}=\mathcal{D}_{-v}E \ket{+v}=E\ket{g_1}. \label{same}
\eeq

\subsection{Finding $\ket{+v}$}\label{solving+v}
To set up our perturbative expansion, let us expand the Hamiltonian about a harmonic oscillator Hamiltonian $H_0$
\beq\label{reqmh'}
H'=H_0+\frac{x_0}{v}H_1 + \left(\frac{x_0}{v}\right)^2 H_2,
\eeq
where
\beq\label{qmh0h1h2}
H_0=\frac{p^2}{2m}+\frac{m \omega^2x_0^2}{2} \frac{x^2}{x_0^2},\quad H_1=\frac{m \omega^2x_0^2}{2}\frac{x^3}{x^3_0},\quad H_2=\frac{m \omega^2x_0^2}{8} \frac{x^4}{x_0^4}\hsp
\omega = \sqrt{\frac{8k v^2}{m}}.
\eeq
Here $x_0$ is an arbitrary unit of length, which we introduced to make $x_0/v$ dimensionless.  We imagine that $x_0/v$ is small while $x_0$, $m$ and $\omega$ are fixed, and we will calculate states and energies as a power series in $x_0/v$.   In this approximation the two wells are well-separated and are treated as two decoupled anharmonic oscillators.  Here $H_0$, $H_1$ and $H_2$ are the leading order, 1st order and 2nd order terms in the expansion of the Hamiltonian, and there are no higher order terms.

Let $\ket{0}$ be the ground state of $H_0$
\beq
H_0\ket{0}=E_0\ket{0}. \label{e0}
\eeq
We will expand $g_1$ as
\beq
\ket{g_1}=\ket{0}+\frac{x_0}{v}\ket{1}+ \left(\frac{x_0}{v}\right)^2\ket{2}+O \left(\left(\frac{x_0}{v}\right)^3\right)
\eeq
and its energy as
\beq
E=E_0+\frac{x_0}{v}E_1 + \left(\frac{x_0}{v}\right)^2 E_2+\mathcal{O}\left(\left(\frac{x_0}{v}\right)^3\right). \label{enex}
\eeq
Then to second order in perturbation theory, the Schrodinger equation is
\bea\label{qmscheq}
&&\left(H_0+\frac{x_0}{v}H_1 + \left(\frac{x_0}{v}\right)^2 H_2\right)\left(\ket{0}+\frac{x_0}{v}\ket{1}+ \left(\frac{x_0}{v}\right)^2\ket{2}+\mathcal{O}\left(\left(\frac{x_0}{v}\right)^3\right)\right)\nonumber\\
&=&\left(E_0+\frac{x_0}{v}E_1 + \left(\frac{x_0}{v}\right)^2 E_2+\mathcal{O}\left(\left(\frac{x_0}{v}\right)^3\right)\right)\left(\ket{0}+\frac{x_0}{v}\ket{1}+ \left(\frac{x_0}{v}\right)^2\ket{2}+\mathcal{O}\left(\left(\frac{x_0}{v}\right)^3\right)\right).\nonumber\\
\eea
To simplify calculations below, we fix unit normalization not on $\ket{g_1}$ but rather on $\ket{0}$, which fixes the normalization of $\ket{g_1}$ so long as each other term $\ket{n}$ is orthogonal to $\ket{0}$
\beq
\braket{0}{0}=1,\quad \braket{0}{1}=\braket{0}{2}=...=\braket{0}{n}=0. \label{orth}
\eeq

Now our task is to express $\ket{1}$ and $\ket{2}$ in terms of operators acting on $\ket{0}$. Let us define the Heisenberg operators $a$ and $a^\dagger$ as usual by
\beq
a=\sqrt{\frac{m\omega}{2}}\left(x+\frac{ip}{m\omega}\right), \quad  a^\dagger=\sqrt{\frac{m\omega}{2}}\left(x-\frac{ip}{m\omega}\right),
\eeq
then (\ref{qmh0h1h2}) becomes
\beq\label{qmaadagger}
H_0=\omega \left(a^\dagger a +\tfrac{1}{2}\right),\quad H_1=\sqrt{\frac{\omega}{32mx_0^2}}(a+a^\dagger)^3,\quad H_2=\frac{1}{32 m x_0^2}(a+a^\dagger)^4.
\eeq

\subsubsection*{0th order}
 First, the terms of order unity in (\ref{qmscheq}) reproduce Eq.~(\ref{e0}) with the unperturbed energy
 \begin{equation}
 E_0=\frac{1}{2}\omega .
 \end{equation}
 
\subsubsection*{1st order}
 At first order we consider the terms of $O\left(\frac{x_0}{v}\right)$ in (\ref{qmscheq}), 
\begin{eqnarray}
 H_0\ket{1}+H_1\ket{0}=E_1\ket{0}+E_0\ket{1}.
\end{eqnarray}
In terms of $a$ and $a^\dagger$ in (\ref{qmaadagger}) this is,
\beq\label{qm1st}
\omega \left(a^\dagger a +\tfrac{1}{2}\right)\ket{1}+\frac{1}{4x_0}\sqrt{\frac{\omega}{2m}}(a+a^\dagger)^3\ket{0}=E_1\ket{0}+E_0\ket{1}.
\eeq
Solving (\ref{qm1st}), subject to the conditions (\ref{orth}), we find
\beq
E_1=0, \quad \ket{1}= -\frac{1}{12x_0\sqrt{2m \omega}}\left(9 a^\dagger +a^\dagger a^\dagger a^\dagger \right)\ket{0}. \label{first}
\eeq
 
\subsubsection*{2nd order}
Finally, we consider the terms of $O\left(\left(\frac{x_0}{v}\right)^2\right)$ in (\ref{qmscheq}), 
\begin{equation}
H_0\ket{2}+H_1\ket{1}+H_2\ket{0}=E_0\ket{2}+E_1\ket{1}+E_2\ket{0}.
\end{equation}
In terms of $a$ and $a^\dagger$,
\beq\label{qm2nd}
\omega \left(a^\dagger a +\tfrac{1}{2}\right)\ket{2}+\frac{1}{4x_0}\sqrt{\frac{\omega}{2m}}(a+a^\dagger)^3\ket{1}+\frac{1}{32 m x_0^2}(a+a^\dagger)^4\ket{0}=E_0\ket{2}+E_1\ket{1}+E_2\ket{0}.
\eeq
Solving (\ref{orth}) and (\ref{qm2nd}) we obtain
\beq
E_2=-\frac{1}{4 m x_0^2}, \quad \ket{2}=\frac{1}{576m \omega x_0^2}\left(189{a^\dagger}^2+27{a^\dagger}^4+{a^\dagger}^6\right)\ket{0}. \label{secondqm}
\eeq 

\subsubsection*{Putting It All Together}
Thus, to the 2nd order, the energy of the ground state is
\beq\label{energyofdoublewell}
E=E_0+\frac{x_0}{v}E_1 + \left(\frac{x_0}{v}\right)^2 E_2=\frac{\omega}{2}-\frac{1}{4 m v^2}
\eeq
and the ground state $\ket{g_1}$ is 
\beq
\ket{g_1}=\left[1-\frac{1}{v}\frac{1}{\sqrt{2m \omega}}\left(\frac{3}{4}a^\dagger +\frac{1}{12}{a^\dagger}^3\right)+\frac{1}{v^2}\left(\frac{21}{64m\omega}{a^\dagger}^2+\frac{3}{64m\omega}{a^\dagger}^4+\frac{1}{576m\omega}{a^\dagger}^6\right)\right]\ket{0}.\nonumber\\
\eeq

We can represent this state by a wave function. Recall that the $n$th excited state of the harmonic oscillator is $\frac{(a^\dagger)^n}{\sqrt {n!}}\ket{0}$, and
\beq
\bra{x}\frac{(a^\dagger)^n}{\sqrt {n!}}\ket{0}=\frac{1}{\sqrt{2^n n!}}\left(\frac{m \omega}{\pi}\right)^{1/4}e^{-\frac{m\omega x^2}{2}}H_n(\sqrt{m\omega}x),
\eeq
where $H_n$ is the $n$th Hermite polynomial. The wave function of $\ket{g_1}$ is
\bea\label{wavefunc+v}
\phi_{g_1}(x)&=&\braket{x}{g_1}\nonumber\\
&=&\frac{1}{\pi^{1 / 4}} e^{-\frac{1}{2} m \omega x^{2}}\Bigg[m^{\frac{1}{4}} \omega^{\frac{1}{4}}-\frac{1}{v}\left(\frac{1}{2} m^{\frac{1}{4}} \omega^{\frac{1}{4}} x+\frac{1}{6} m^{\frac{5}{4}} \omega^{\frac{5}{4}} x^{3}\right)\nonumber\\
&&+\frac{1}{v^{2}}\left(-\frac{41}{192} m^{-\frac{3}{4}} \omega^{-\frac{3}{4}}+\frac{1}{4} m^{\frac{1}{4}} \omega^{\frac{1}{4}} x^{2}+\frac{1}{12} m^{\frac{5}{4}} \omega^{\frac{5}{4}} x^{4}+\frac{1}{72} m^{\frac{9}{4}} \omega^{\frac{9}{4}} x^{6}\right)\Bigg]
\eea
which is normalized such that
\beq
\int dx \phi_{g_1}^2(x)=1+\frac{29}{96 m \omega v^{2}}+\frac{4981}{18432 m^{2} \omega^{2} v^{4}}.
\eeq
As a check, one can substitute (\ref{wavefunc+v}) and (\ref{energyofdoublewell}) into the full Schrodinger equation,
\bea
&&H'\phi_{g_1}(x)-E\phi_{g_1}(x)\nonumber\\
&=&\frac{1}{m \pi^{1 / 4}} e^{-\frac{1}{2} m \omega x^{2}} \Bigg[\frac{1}{v^{3}}\left(-\frac{m^{\frac{1}{4}} \omega^{\frac{1}{4}} x}{8}-\frac{19 m^{\frac{5}{4}} \omega^{\frac{5}{4}} x^{3}}{128}+\frac{m^{\frac{9}{4}} \omega^{\frac{9}{4}} x^{5}}{16}+\frac{m^{\frac{13}{4}} \omega^{\frac{13}{4}} x^{7}}{48}+\frac{m^{\frac{17}{4}} \omega^{\frac{17}{4}} x^{9}}{144}\right)\nonumber\\ 
&&\frac{1}{v^4}\left(-\frac{41 m^{-\frac{3}{4}} \omega^{-\frac{3}{4}}}{768}+\frac{m^{\frac{1}{4}} \omega^{\frac{1}{4}} x^{2}}{16}-\frac{3 m^{\frac{5}{4}} \omega^{\frac{5}{4}} x^{4}}{512}+\frac{5 m^{\frac{9}{4}} \omega^{\frac{9}{4}} x^{6}}{144}+\frac{m^{\frac{13}{4}} \omega^{\frac{13}{4}} x^{8}}{96}+\frac{m^{\frac{17}{4}} \omega^{\frac{17}{4}} x^{10}}{576}\right)\Bigg]\nonumber\\
&=&\mathcal{O}\left(\left(\frac{1}{v}\right)^3\right).
\eea
As expected, the three leading terms have all canceled, leaving only higher order terms.  Of course these higher order terms are large when $x$ is far from the bottom of the well centered at $x=0$, as they will be at any order in perturbation theory.

One may obtain the ground state wave function $\ket{+v}$ of the original Hamiltonian $H$ using the shift operator
\bea
\phi_{+v}(x)&=&\mathcal{D}_{+v} \phi_{g_1}(x) \nonumber\\
&=&\frac{1}{\pi^{1 / 4}} e^{-\frac{1}{2} m \omega (x-v)^{2}}\Bigg[m^{\frac{1}{4}} \omega^{\frac{1}{4}}-\frac{1}{v}\left(\frac{1}{2} m^{\frac{1}{4}} \omega^{\frac{1}{4}} (x-v)+\frac{1}{6} m^{\frac{5}{4}} \omega^{\frac{5}{4}} (x-v)^{3}\right)\nonumber\\
&&+\frac{1}{v^{2}}\left(-\frac{41}{192} m^{-\frac{3}{4}} \omega^{-\frac{3}{4}}+\frac{1}{4} m^{\frac{1}{4}} \omega^{\frac{1}{4}} (x-v)^{2}+\frac{1}{12} m^{\frac{5}{4}} \omega^{\frac{5}{4}} (x-v)^{4}+\frac{1}{72} m^{\frac{9}{4}} \omega^{\frac{9}{4}} (x-v)^{6}\right)\Bigg].\nonumber\\
\eea
Note that this shift mixes the orders in our perturbative expansion, and so this wave function is not a solution to the original Schrodinger equation beyond the leading order.  In fact, we have not defined a perturbative expansion of the original Hamiltonian $H$.  

To obtain the corresponding quantities for the state $\ket{-v}$, one need only change the sign of $v$ in the formulas above.

\section{Constructing A Perturbative $\phi^4$ Ground State} \label{qft}

We are interested in the 1+1 dimensional quantum field theory of a real scalar field $\phi$ with a double-well $\phi^4$ interaction
\beq\label{hphi4}
H=\int dx :\frac{1}{2}\left(\partial_x\phi\right)^2 + \frac{1}{2}\pi^2 +\frac{\lambda}{4}\left(\phi^2-v^2\right)^2:
\eeq
where the normal ordering is defined below and $\pi$ is the momentum conjugate to $\phi$.  We will also define a mass
\beq
m=v\sqrt{2\lambda}
\eeq
which corresponds to the mass of the field when expanded about the bottom of either well.  In 1+1 dimensions $\phi$ is dimensionless, $\lambda$ has dimensions of $m^2$ and $v$ is dimensionless. We will consider a perturbative expansion in $1/v$.  This theory has two degenerate ground states of energy $E$, centered near $\phi=v$ and $\phi=-v$, which we will call $\ket{+v}$ and $\ket{-v}$ respectively.   In this section we will construct $\ket{+v}$.



Although $\phi$ is not a free field, in the Schrodinger picture we can Fourier transform $\phi$ and $\pi$ to define $a$ and $a^\dag$
\begin{eqnarray}
\phi(x)&=&\int \frac{d p}{2 \pi} \frac{1}{\sqrt{2 \omega_{p}}}\left(a_{p}+a_{-p}^{\dagger}\right) e^{i p x},\nonumber\\
\pi(x)&=&\int \frac{d p}{2 \pi}(-i) \sqrt{\frac{\omega_{p}}{2}}\left(a_{p}-a_{-p}^{\dagger}\right) e^{i p x}
\end{eqnarray}
where
\beq
\omega_p=\sqrt{m^2+p^2}.
\eeq
Now the canonical commutation relations
\beq
[\phi(x),\pi(y)]=i\delta(x-y) \label{phipi}
\eeq
imply that $a$ and $a^\dag$ obey the Heisenberg algebra
\beq
[a_p,a^\dag_q]=2\pi\delta(p-q)
\eeq
even in the interacting theory.  Our normal ordering prescription is that all $a^\dag$ be placed to the left of all $a$.


\subsection{The Displacement Operator $\mathcal{D}$}
The displacement operator
\begin{equation}
\mathcal{D}_\alpha=e^{-i\alpha\int dx \pi(x)}=e^{\alpha \sqrt{m/2}\left(a_0^\dagger-a_0\right)}
\end{equation}
shifts the value of $\phi$ by $\alpha$ 
\begin{equation}
\left[\phi,\mathcal{D}_\alpha\right]=\alpha \mathcal{D}_\alpha.
\end{equation}
Also it commutes with normal ordering.   

Applying it to the Hamiltonian $H$ one finds the shifted Hamiltonian
\beq
H\p=\mathcal{D}_{-v}H \mathcal{D}_{-v}^{-1}
\eeq
given by
\begin{equation}\label{h'phi4}
H'=\int dx : \frac{1}{2}\left(\partial_x\phi\right)^2 + \frac{1}{2}\pi^2 +\frac{1}{2}m^2\phi^2+\frac{1}{v}\frac{m^2}{2}\phi^3+\frac{1}{v^2}\frac{m^2}{8}\phi^4 :.
\end{equation}

\subsection{Finding the Operator Perturbatively}

We denote the ground state of (\ref{h'phi4}) as $\ket{g_1}$. So we have the relation
\begin{equation}
\ket{+v}=\mathcal{D}_{+v}\ket{g_1}
\end{equation}
and Eq.~(\ref{same}) applies as written to the quantum field theory case, showing that $\ket{g_1}$ has energy $E$.  Again our goal is to write $\ket{g_1}$ in terms of the free ground state.


First we perturbatively expand the Hamiltonian (\ref{h'phi4}),
\beq\label{H'}
H'=H_0+\frac{1}{v}H_1+\frac{1}{v^2}H_2
\eeq
where
\beq
H_0=\int dx {:\frac{1}{2}{\phi'}^2+\frac{1}{2}\pi^2+\frac{1}{2}m^2\phi^2:}\hsp
H_1={\frac{1}{v}\int dx \frac{m^2}{2}:\phi^3:}\hsp 
H_2={\frac{1}{v^2}\int dx \frac{m^2}{8}:\phi^4:}.
\eeq
$H_0$ is the Hamiltonian of the free, massive Klein-Gordon theory, and the next two terms are perturbations of order $O\left(\frac{1}{v}\right)$ and $O\left(\frac{1}{v^2}\right)$. 

Expanding the ground state
\beq
\ket{g_1}=\left(\ket{0}+\frac{1}{v}\ket{1}+\frac{1}{v^2}\ket{2} \right)+\mathcal{O}\left(\frac{1}{v^3}\right)
\eeq
subject to (\ref{orth}) and the energy as in Eq.~(\ref{enex}) we obtain the eigenvalue equation for $\ket{g_1}$
\begin{eqnarray}\label{schrodinger_equation_to_O2}
&&\left( H_0+\frac{1}{v}H_1+ \frac{1}{v^2}H_2 \right)\left(\ket{0}+\frac{1}{v}\ket{1}+\frac{1}{v^2}\ket{2}+\mathcal{O}\left(\frac{1}{v^3}\right) \right)\nonumber\\
&=&\left(E_0+\frac{1}{v}E_1+\frac{1}{v^2}E_2+\mathcal{O}\left(\frac{1}{v^3}\right)\right)\left(\ket{0}+\frac{1}{v}\ket{1}+\frac{1}{v^2}\ket{2}+\mathcal{O}\left(\frac{1}{v^3}\right)\right).
\end{eqnarray}

\subsubsection*{0th order}
The leading terms in (\ref{schrodinger_equation_to_O2}) are
\begin{equation}
H_0 \ket{0}=E_0 \ket{0} \label{free}
\end{equation}
so that $\ket{0}$ is the free ground state and
\beq
E_0=0. \label{tired}
\eeq
Note that this is superficially different from the energy in the quantum mechanical case, because we have normal ordered our Hamiltonian.  Of course had we normal ordered in that case we would also have obtained a vanishing value of $E_0$.

\subsubsection*{1st order}
The first order terms in (\ref{schrodinger_equation_to_O2}) are
\begin{eqnarray}\label{O1eq}
H_0\ket{1}+H_1\ket{0}=E_1\ket{0}.
\end{eqnarray}
Using the normal ordering in $H_1$ and the fact that 
\beq
a_p\ket{0}=0
\eeq
as a result of (\ref{free}) and (\ref{tired}), this can be written explicitly as
\begin{equation}\label{H0|1>}
H_0\ket{1}=E_1\ket{0}-\frac{m^2}{2}\int \frac{dp_1 dp_2}{(2\pi)^2}\frac{a_{p_1}^\dagger a_{p_2}^\dagger a_{-p_1-p_2}^\dagger}{\sqrt{8 \omega_{p_1}\omega_{p_2}\omega_{p_1+p_2}}} \ket{0}.
\end{equation}

Applying $\langle 0|$ to (\ref{H0|1>}), together with the condition (\ref{orth}), we find
\begin{equation}
E_1=0
\end{equation}
and so then (\ref{H0|1>}) is solved by inverting $H_0$, yielding a sum over frequencies in the denominator
\begin{equation}
\ket{1}=-\frac{m^2}{2}\int \frac{dp_1 dp_2}{(2\pi)^2}\frac{1}{\sqrt{8 \omega_{p_1}\omega_{p_2}\omega_{p_1+p_2}}}\frac{a_{p_1}^\dagger a_{p_2}^\dagger a_{-p_1-p_2}^\dagger}{\left(\omega_{p_1}+\omega_{p_2}+\omega_{p_1+p_2}\right)} \ket{0}. \label{firstqft}
\end{equation}
This looks similar to (\ref{first}), except that the $a^\dag$ term is missing.  This again is the result of the normal ordering.

\subsubsection*{2nd order}
Finally we arrive at the $O\left(\frac{1}{v^2}\right)$ terms in (\ref{schrodinger_equation_to_O2}), 
\begin{equation}
H_0\ket{2}+H_1\ket{1}+H_2\ket{0}=E_2\ket{0}. \label{seceq}
\end{equation}
Applying $\langle 0|$ on the left, the first term vanishes because $E_0=0$ and the $H_2$ term vanishes because it is normal ordered.  Using the normalization of $|0\rangle$ we obtain
\beq
E_2=\langle 0|H_1\ket{1}.
\eeq
Using (\ref{firstqft}) we find the ground state energy at this order
\beq
E_2=\int dx {\mathcal{E}}_2(x)
\eeq
\beq
{\mathcal{E}}_2(x)
=-\frac{3}{16}m^4 \int \frac{dp_1}{2\pi}\frac{dp_2}{2\pi}\frac{1}{\omega_{p_1}\omega_{p_2}\omega_{p_1+p_2}} \frac{1}{(\omega_{p_1}+\omega_{p_2}+\omega_{p_1+p_2})}.
\eeq
The energy is infinite, however the energy density $\mathcal{E}_2$ is finite and constant. To the 2nd order, the energy density of the $\phi^4$ vacuum state is
\begin{eqnarray}
\mathcal{E}&=&0+\frac{1}{v}\mathcal{E}_1+\frac{1}{v^2}{\mathcal{E}_2}\nonumber\\
&=&-\frac{1}{v^2}\frac{3}{16}m^4 \int \frac{dp_1}{2\pi}\frac{dp_2}{2\pi}\frac{1}{\omega_{p_1}\omega_{p_2}\omega_{p_1+p_2}} \frac{1}{(\omega_{p_1}+\omega_{p_2}+\omega_{p_1+p_2})} \nonumber\\
&\approx&-0.022 \lambda.
\end{eqnarray}

Now we know everything in Eq.~(\ref{seceq}) except for $\ket{2}$.  This can again be found by inverting $H_0$, providing another sum over frequencies in the denominator
\begin{eqnarray}
\ket{2}&=&\bigg [\frac{9}{4}m^4\int \frac{dp_1 dp_2}{(2\pi)^2} \frac{1}{8\omega_{p_1}^2 \omega_{p_2} \omega_{p_1+p_2}} \frac{a_{p_1}^\dagger a_{-p_1}^\dagger}{(\omega_{p_1}+\omega_{p_2}+\omega_{p_1+p_2})}\nonumber\\
&&+\int \frac{dp_1 dp_2 dp_3}{(2\pi)^3} \frac{1}{\sqrt{16 \omega_{p_1} \omega_{p_2} \omega_{p_3} \omega_{p_1+p_2+p_3}}}\frac{a_{p_1}^\dagger a_{p_2}^\dagger a_{p_3}^\dagger a_{-p_1-p_2-p_3}^\dagger}{(\omega_{p_1}+\omega_{p_2}+\omega_{p_3}+\omega_{p_1+p_2+p_3})}\nonumber\\
&&\times \left({\frac{9}{4}m^4\frac{1}{2\omega_{p_1+p_2}(\omega_{p_1+p_2}+\omega_{p_3}+\omega_{p_1+p_2+p_3})}-\frac{m^2}{8}} \right) \nonumber\\
&&+\int \frac{dp_1 dp_2 dp_3 dp_4}{(2\pi)^4} \frac{m^4}{4}\frac{1}{\omega_{p_1}+\omega_{p_2}+\omega_{p_1+p_2}+\omega_{p_3}+\omega_{p_4}+\omega_{p_3+p_4}}\nonumber\\
&&\times \frac{1}{\sqrt{64 \omega_{p_1}\omega_{p_2}\omega_{p_1+p_2}\omega_{p_3}\omega_{p_4}\omega_{p_3+p_4}}} \frac{a_{p_1}^\dagger a_{p_2}^\dagger a_{-p_1-p_2}^\dagger a_{p_3}^\dagger a_{p_4}^\dagger a_{-p_3-p_4}^\dagger}{(\omega_{p_3}+\omega_{p_4}+\omega_{p_3+p_4})}\bigg] \ket{0}.
\end{eqnarray}
Note that it resembles the corresponding expression in quantum mechanics (\ref{secondqm}).  The two particle, four particle and six particle components have just the same origin in both cases.

Now we can write $\ket{g_1}$ to 2nd order
\begin{eqnarray}
\ket{g_1}&=&\Bigg \{ 1-\frac{1}{v}\frac{m^2}{2}\int \frac{dp_1 dp_2}{(2\pi)^2}\frac{1}{\sqrt{8 \omega_{p_1}\omega_{p_2}\omega_{p_1+p_2}}}\frac{a_{p_1}^\dagger a_{p_2}^\dagger a_{-p_1-p_2}^\dagger}{\left(\omega_{p_1}+\omega_{p_2}+\omega_{p_1+p_2}\right)}\nonumber\\
&&+\frac{1}{v^2}\bigg [\frac{9}{4}m^4\int \frac{dp_1 dp_2}{(2\pi)^2} \frac{1}{8\omega_{p_1}^2 \omega_{p_2} \omega_{p_1+p_2}} \frac{a_{p_1}^\dagger a_{-p_1}^\dagger}{(\omega_{p_1}+\omega_{p_2}+\omega_{p_1+p_2})}\nonumber\\
&&+\int \frac{dp_1 dp_2 dp_3}{(2\pi)^3} \frac{1}{\sqrt{16 \omega_{p_1} \omega_{p_2} \omega_{p_3} \omega_{p_1+p_2+p_3}}}\frac{a_{p_1}^\dagger a_{p_2}^\dagger a_{p_3}^\dagger a_{-p_1-p_2-p_3}^\dagger}{(\omega_{p_1}+\omega_{p_2}+\omega_{p_3}+\omega_{p_1+p_2+p_3})}\nonumber\\
&&\times \left({\frac{9}{4}m^4\frac{1}{2\omega_{p_1+p_2}(\omega_{p_1+p_2}+\omega_{p_3}+\omega_{p_1+p_2+p_3})}-\frac{m^2}{8}} \right) \nonumber\\
&&+\int \frac{dp_1 dp_2 dp_3 dp_4}{(2\pi)^4} \frac{m^4}{4}\frac{1}{\omega_{p_1}+\omega_{p_2}+\omega_{p_1+p_2}+\omega_{p_3}+\omega_{p_4}+\omega_{p_3+p_4}}\nonumber\\
&&\times \frac{1}{\sqrt{64 \omega_{p_1}\omega_{p_2}\omega_{p_1+p_2}\omega_{p_3}\omega_{p_4}\omega_{p_3+p_4}}} \frac{a_{p_1}^\dagger a_{p_2}^\dagger a_{-p_1-p_2}^\dagger a_{p_3}^\dagger a_{p_4}^\dagger a_{-p_3-p_4}^\dagger}{(\omega_{p_3}+\omega_{p_4}+\omega_{p_3+p_4})}\bigg] \Bigg \} \ket{0}. \label{g1eq}
\end{eqnarray}
Let us denote the operator in the curly bracket by $\mathcal{O}\p$
\beq
\ket{g_1}=\mathcal{O}\p \ket{0}.
\eeq
Let the operator which takes the free vacuum $\ket{0}$ to the $\phi^4$ vacuum $\ket{+v}$ be $\mathcal{O}_1$
\beq
\ket{+v}=\mathcal{O}_1\ket{0}
\eeq
then
\beq
\mathcal{O}_1=\mathcal{D}_{+v} \mathcal{O}\p.
\eeq

\section{Interpolating Operator} \label{interp}
To obtain $\ket{-v}$, the other ground state, we perform exactly the same manipulations but with the sign of $v$ flipped.   The resulting state $\ket{g_1}$ is identical to that in Eq.~(\ref{g1eq}) except each $v$ is replaced by $-v$.  Let the corresponding operator, corresponding to $\mathcal{O}\p$ with $v\rightarrow -v$, be $\mathcal{O}\p{}\p$ and the operator that changes $\ket{0}$ to $\ket{-v}$ be $\mathcal{O}_2$
\beq
\mathcal{O}_2=\mathcal{D}_{-v} \mathcal{O}\p{}\p.
\eeq

The operator $\mathcal{O}$ that changes $\ket{+v}$ to $\ket{-v}$
\beq
\ket{-v}=\mathcal{O}\ket{+v}
\eeq
is then
\begin{eqnarray}
\mathcal{O}&=&\mathcal{O}_2\mathcal{O}_1^{-1}\nonumber\\
&=&\mathcal{D}_{-v} \mathcal{O}'' \left(\mathcal{D}_{+v} \mathcal{O}'\right)^{-1}\nonumber\\
&=&\mathcal{D}_{-v} \mathcal{O}''\mathcal{O}'^{-1} \mathcal{D}_{-v}.
\end{eqnarray}
Let us define the shorthand notation
\begin{eqnarray}
\mathcal{O}'&=&1-\frac{1}{v}\mathcal{A}+\frac{1}{v^2}\mathcal{B},\\
\mathcal{O}''&=&1+\frac{1}{v}\mathcal{A}+\frac{1}{v^2}\mathcal{B}\nonumber
\end{eqnarray}
where
\begin{equation}
\mathcal{A}=\frac{m^2}{2}\int \frac{dp_1 dp_2}{(2\pi)^2}\frac{1}{\sqrt{8 \omega_{p_1}\omega_{p_2}\omega_{p_1+p_2}}}\frac{a_{p_1}^\dagger a_{p_2}^\dagger a_{-p_1-p_2}^\dagger}{\left(\omega_{p_1}+\omega_{p_2}+\omega_{p_1+p_2}\right)},
\end{equation}
\begin{eqnarray}
\mathcal{B}&=&\frac{9}{4}m^4\int \frac{dp_1 dp_2}{(2\pi)^2} \frac{1}{8\omega_{p_1}^2 \omega_{p_2} \omega_{p_1+p_2}} \frac{a_{p_1}^\dagger a_{-p_1}^\dagger}{(\omega_{p_1}+\omega_{p_2}+\omega_{p_1+p_2})}\nonumber\\
&&+\int \frac{dp_1 dp_2 dp_3}{(2\pi)^3} \frac{1}{\sqrt{16 \omega_{p_1} \omega_{p_2} \omega_{p_3} \omega_{p_1+p_2+p_3}}}\frac{a_{p_1}^\dagger a_{p_2}^\dagger a_{p_3}^\dagger a_{-p_1-p_2-p_3}^\dagger}{(\omega_{p_1}+\omega_{p_2}+\omega_{p_3}+\omega_{p_1+p_2+p_3})}\nonumber\\
&&\times \left({\frac{9}{4}m^4\frac{1}{2\omega_{p_1+p_2}(\omega_{p_1+p_2}+\omega_{p_3}+\omega_{p_1+p_2+p_3})}-\frac{m^2}{8}} \right) \nonumber\\
&&+\int \frac{dp_1 dp_2 dp_3 dp_4}{(2\pi)^4} \frac{m^4}{4}\frac{1}{\omega_{p_1}+\omega_{p_2}+\omega_{p_1+p_2}+\omega_{p_3}+\omega_{p_4}+\omega_{p_3+p_4}}\nonumber\\
&&\times \frac{1}{\sqrt{64 \omega_{p_1}\omega_{p_2}\omega_{p_1+p_2}\omega_{p_3}\omega_{p_4}\omega_{p_3+p_4}}} \frac{a_{p_1}^\dagger a_{p_2}^\dagger a_{-p_1-p_2}^\dagger a_{p_3}^\dagger a_{p_4}^\dagger a_{-p_3-p_4}^\dagger}{(\omega_{p_3}+\omega_{p_4}+\omega_{p_3+p_4})}.
\end{eqnarray}
Then,
\begin{eqnarray}
\mathcal{O}'^{-1}&=&\left(1-\frac{1}{v}\mathcal{A}+\frac{1}{v^2}\mathcal{B}\right)^{-1}\nonumber\\
&=&1+\frac{1}{v}\mathcal{A}+\frac{1}{v^2}\left(\mathcal{A}^2-\mathcal{B}\right).
\end{eqnarray}
Finally,
\begin{eqnarray}
\mathcal{O}&=&\mathcal{D}_{-v} \left(1+\frac{1}{v}\mathcal{A}+\frac{1}{v^2}\mathcal{B}\right) \left(1+\frac{1}{v}\mathcal{A}+\frac{1}{v^2}\left(\mathcal{A}^2-\mathcal{B}\right)\right)\mathcal{D}_{-v}\nonumber\\
&=&\mathcal{D}_{-v}\left[1+\frac{2}{v}\mathcal{A}+\frac{2}{v^2}\mathcal{A}^2\right]\mathcal{D}_{-v}\nonumber\\
&=&e^{-v \sqrt{\frac{m}{2}}\left(a_0^\dagger-a_0\right)} \Bigg[1+\frac{m^2}{v}\int \frac{dp_1 dp_2}{(2\pi)^2}\frac{1}{\sqrt{8 \omega_{p_1}\omega_{p_2}\omega_{p_1+p_2}}}\frac{a_{p_1}^\dagger a_{p_2}^\dagger a_{-p_1-p_2}^\dagger}{\left(\omega_{p_1}+\omega_{p_2}+\omega_{p_1+p_2}\right)}\nonumber\\
&&+\frac{m^4}{2v^2}\left(\int \frac{dp_1 dp_2}{(2\pi)^2}\frac{1}{\sqrt{8 \omega_{p_1}\omega_{p_2}\omega_{p_1+p_2}}}\frac{a_{p_1}^\dagger a_{p_2}^\dagger a_{-p_1-p_2}^\dagger}{\left(\omega_{p_1}+\omega_{p_2}+\omega_{p_1+p_2}\right)}\right)^2\Bigg] e^{-v \sqrt{\frac{m}{2}}\left(a_0^\dagger-a_0\right)}.
\end{eqnarray}
At this order, there is no dependence on $\mathcal{B}$.

The main motivation of the present work is to provide necessary formulas for the calculation of higher corrections to the $\phi^4$ kink.  If the center of the kink is at $x_0$ and the kink form factor, in the sense of Ref.~\cite{taylor78}, interpolates between $-v$ at $x\rightarrow -\infty$ and $+v$ at $x\rightarrow\infty$, then we expect that in the suitably defined limit $x_0\rightarrow\infty$ the operator which creates the kink will reduce to $\mathcal{O}$.

\section{Remarks}

In this note we have performed a simple exercise in perturbative quantum field theory, finding the ground states of the $\phi^4$ double-well theory and their energies.   Armed with this perturbative expression for the ground state, we may now extend the calculation of kink state and mass in Ref.~\cite{mekink} beyond the leading quantum corrections.  We wish to do this because our eventual goal, in the supersymmetric case, is to attempt to construct the kink at all orders but we would like the first terms in the perturbation theory as a starting point.

Another possible application of this ground state is in Higgs physics.  The Higgs field is of this form, except that it is complex and also lives in 3+1 dimensions.  However the radial mode, near the minimum, is governed by the same potential.  A quantum description of this state may help one to study the following phenomena.

A superconductor can repel a magnetic field.  It does this by letting its electrons travel in circles, which creates a counter magnetic field.  This costs energy, and so changes the potential energy as a function of the Cooper pair condensate.  The field equations imply that this energy must be minimized and so, in the presence of an external magnetic field, it is well known that the Cooper pair condensate is reduced.  

Exactly the same argument may be applied to the Standard Model Higgs field.  It creates mass for, for example, the top quark.  But this mass costs the Higgs field energy, and so in the minimum energy field configuration, in the presence of a top quark, the Higgs condensate will be reduced.   This reduction is quite localized about the top quark, which anyway does not live long.  However, in a proposed 100 TeV collider, one expects that there will be a larger region where a number of particles are given masses by the Higgs field.  If this energy density is of order the Higgs potential, then the backreaction on the Higgs field, reducing its value and the particle masses, may be considerable.   This would be new physics, but simply calculated in perturbation theory with this potential coupled via a Yukawa coupling to fermions.  This is the same coupling that we will need to add anyway to study the supersymmetric kink.

\section* {Acknowledgement}

\noindent
JE is supported by the CAS Key Research Program of Frontier Sciences grant QYZDY-SSW-SLH006 and the NSFC MianShang grants 11875296 and 11675223.  JE also thanks the Recruitment Program of High-end Foreign Experts for support.



\end{document}